# The generalised Blasius correlation for turbulent flow past flat plates

Trinh, Khanh Tuoc

*K.T.Trinh@massey.ac.nz*

---

## Abstract

This paper presents a theoretical derivation of the empirical Blasius power law correlation for the friction factor. The coefficients in this correlation are shown to be dependent on the Reynolds number. Published experimental data is well correlated.

## 1 Introduction

The velocity profiles in turbulent flow can be described in terms of a log-law as well as a power law (Schlichting, 1960). The log-law first proposed by Prandtl (1935) and later justified more formally by Millikan (1938) by similarity arguments is usually written as

$$U^+ = \frac{1}{\kappa}\ln y^+ + B \tag{1}$$

where the velocity $U^+ = U/u_*$ and the normal distance from the wall $y^+ = yu_*\rho/\mu = yu_*/\nu$ have been normalised with the friction velocity $u_* = \sqrt{\tau_w/\rho}$ and the fluid viscosity $\mu$. The parameters $\kappa$ called Karman's constant and B vary slightly with experimental data and especially with flow geometry but are often quoted with the canonical values *0.4* and 2.5 respectively as first reported by Nikuradse (1932).

In pipe flow, the friction factor

$$f = \frac{2\tau_w}{\rho V^2} \qquad (2)$$

is usually derived from the velocity $U_m^+$ at the pipe axis where $y^+ = R^+$. Here $V$ is the average flow velocity in the pipe and $R$ is the radius. This derivation (Prandtl, 1935, Karman, 1934) almost coincides with the experimental result of Nikuradse

$$\frac{1}{\sqrt{f}} = 4.0 \log\left(\mathrm{Re}\sqrt{f}\right) - 0.4 \qquad (3)$$

where

$$\mathrm{Re} = \frac{DV\rho}{\mu} \text{ is the Reynolds number} \qquad (4)$$

The thickness of an external boundary layer can, on the other hand, grow indefinitely. Application of the log-law to flow past flat plates e.g. (Schultz-Grunow, 1941) correlates the velocity profile near the wall but does not give the friction factor because the thickness of the turbulent boundary layer in this approach is undefined and the problem is not closed. The traditional method of dealing with this difficulty has been to solve the integral momentum equation numerically, using Prandtl's velocity distribution, to give the boundary layer thickness and the friction factor (Schlichting 1960, p 601).

A second way of describing the velocity profile in turbulent boundary layer flow

$$\frac{U}{U_\infty} = \left(\frac{y}{\delta}\right)^p \qquad (5)$$

originated from the empirical power law relationship introduced by Blasius (1913).

$$f = \frac{0.079}{\mathrm{Re}^{1/4}} \qquad (6)$$

Equation (6) is compatible with a value $p = 1/7$ but applies only up to $\mathrm{Re} \approx 10^5$. Nikuradse (1932) showed that the exponent $p$ in fact decreased as the Reynolds number increased (Schlichting, 1960). Recently, Zagarola, Perry and Smits (1997) have argued from new, more careful measurements of pipe flow data that both the log-law and the power law in the region apply in the inner region but give much more restrictive limits to these so-called overlap regions, particularly for the log-law.

Neglecting, the small but real inconsistencies between these laws and measured velocity profiles, we can derive the Blasius correlation theoretically to avoid the need for numerical integration of the integral momentum equation and account for the varying values of $p$. This is the topic of the present paper.

## 2 Theory

The Blasius correlation may be written in the general power law form

$$f = \frac{\alpha}{\mathrm{Re}_\delta^\beta} \tag{7}$$

Where $\delta$ is the thickness of the boundary layer, yet unknown.

The indices $p$ and $\beta$ are related (Skelland and Sampson, 1973, Trinh, 2010a)

$$p = \frac{\beta}{2-\beta} \tag{8}$$

Equations (2) and (6) give

$$\tau_w = \frac{\alpha}{2}\rho U^{2-\beta}\frac{\nu^\beta}{\delta^\beta} \tag{9}$$

The integral momentum equation for a plate wetted on one side is

$$D(x) = b\rho\int_0^\infty U(U_\infty - U)dy = b\int_0^x \tau_w(x')dx' \tag{10}$$

Introducing the momentum boundary layer thickness defined as

$$\delta_2 U_\infty^2 = \int_0^\delta U(U_\infty - U)dy \tag{11}$$

$$D(x) = b\rho U_\infty^2 \delta_2(x) \tag{12}$$

Substituting for U from equation (5) into (11) and integrating gives

$$\delta_2 = \frac{p}{(p+1)(2p+1)}\delta = \frac{\beta(2-\beta)}{2(2+\beta)}\delta \tag{13}$$

Differentiating equation (12) gives

$$\frac{1}{b}\frac{dD(x)}{dx} = \tau_w = \rho U_\infty^2\frac{d\delta_2}{dx} = \frac{\beta(2-\beta)}{2(2+\beta)}\rho U_\infty^2\frac{d\delta}{dx} \tag{14}$$

and substituting for $\tau_w$ from equation (9)

$$\frac{\tau_w}{\rho U_\infty^2} = \frac{\alpha}{2}\left(\frac{\nu}{U_\infty\delta}\right)^\beta = \frac{\beta(2-\beta)}{2(2+\beta)}\frac{d\delta}{dx} \tag{15}$$

Integrating equation (15) and rearranging gives

$$\delta^{\beta+1} = \frac{\alpha(\beta+2)(\beta+1)}{\beta(2-\beta)}\left(\frac{\nu}{U_\infty}\right)^{\beta} x \tag{16}$$

$$\frac{\delta}{x} = \left[\frac{\alpha(2+\beta)(1+\beta)}{\beta(2-\beta)}\right]^{\frac{1}{\beta+1}}\left(\frac{\nu}{U_\infty x}\right)^{\frac{\beta}{\beta+1}} \tag{17}$$

The issue thus boils down to an estimate of $\beta$. This is done by here by forcing equation (17) a known point in the normalised logarithmic velocity profile described by equation (1). For fully turbulent flow three points are well known (Trinh, 2009): the edge of the wall layer $(\delta_\nu^+ = 64.8, U_\nu^+ = 15.6)$, the edge of the buffer layer $(\delta_b^+ = 30, U_b^+ = 14)$ first described by Karman (op. cit.) and representing a time averaged value of the wall layer thickness (Trinh, 2010c) and the Kolmogorov point $(U_k^+ = y_k^+ = 11.8)$ which is the intersection of equation (1) with the line $U^+ = y^+$. Using the Kolmogorov point gives

$$\frac{11.8}{U_\infty^+} = \left(\frac{11.8}{\delta^+}\right)^p \tag{18}$$

$$p = \frac{\ln\left(11.8/U_\infty^+\right)}{\ln\left(11.8/\delta^+\right)} = \frac{2.461 - \ln U_\infty^+}{2.461 - \ln\delta^+} \tag{19}$$

The unknown boundary layer thickness $\delta$ is eliminated by noting that equation (1) gives

$$U_\infty^+ = 2.5\ln\delta^+ + 5.5 \tag{20}$$

$$\ln\delta^+ = \frac{U_\infty^+ - 5.5}{2.5} \tag{21}$$

Then

$$p = \frac{2.461 - \ln U_\infty^+}{4.668 - U_\infty^+/2.5} = \frac{6.170 - 2.5\ln U_\infty^+}{11.670 - U_\infty^+} \tag{22}$$

Equation (8) can be rearranged as

$$\beta = \frac{2p}{p+1} \tag{23}$$

Substituting for $p$ from equation (23)

$$\beta = \frac{5\ln U_\infty^+ - 12.34}{U_\infty^+ + 2.5\ln U_\infty^+ - 17.840} \tag{24}$$

Using the edge of the wall layer gives

$$\frac{15.6}{U_\infty^+} = \left(\frac{64.8}{\delta^+}\right)^p \qquad (25)$$

$$p = \frac{\ln\left(15.6 \big/ U_\infty^+\right)}{\ln\left(64.8 \big/ \delta^+\right)} = \frac{2.745 - \ln U_\infty^+}{4.17 - \ln \delta^+} \qquad (26)$$

$$\beta = \frac{5 \ln U_\infty^+ - 13.729}{U_\infty^+ + 2.5 \ln U_\infty^+ - 22.793} \qquad (27)$$

Equations (24) and (27) are very similar. We choose to use (27) because the edge of the wall layer is a clear physical position that lies on the log-law.

Substituting for $f$ from equation (2) into (27) gives

$$\beta = \frac{5 \ln\left(\sqrt{2/f}\right) - 13.729}{\sqrt{2/f} + 2.5 \ln\left(\sqrt{2/f}\right) - 22.793} \qquad (28)$$

$$\beta = \frac{2.5 \ln f - 11.996}{\sqrt{2/f} + 1.25 \ln f - 21.926} \qquad (29)$$

The coefficients α is obtained by rearranging equation (7)

$$\alpha = f \,\mathrm{Re}_\delta^\beta = 2 \delta^{+^\beta} U_\infty^{+^{\beta-2}} \qquad (30)$$

Equation (25) can be rearranged as

$$\delta^+ = \frac{64.8}{15.8^{1/p}} U_\infty^{+^{1/p}} = \frac{64.8}{15.8^{2-\beta/\beta}} U_\infty^{+^{2-\beta/\beta}} \qquad (31)$$

Substituting (31) into (30) gives

$$\alpha = \frac{2 \times 64.8^\beta}{15.8^{2-\beta}} \qquad (32)$$

The variations of the factors $\alpha$ and $\beta$ with the friction factor are shown in Figure 1.

The Reynolds number is introduced by noting that equation (17) can be rearranged to give

$$\mathrm{Re}_\delta = \left[\frac{\alpha(2+\beta)(1+\beta)}{\beta(2-\beta)}\right]^{\frac{1}{\beta+1}} \mathrm{Re}_x^{\frac{1}{\beta+1}} \qquad (33)$$

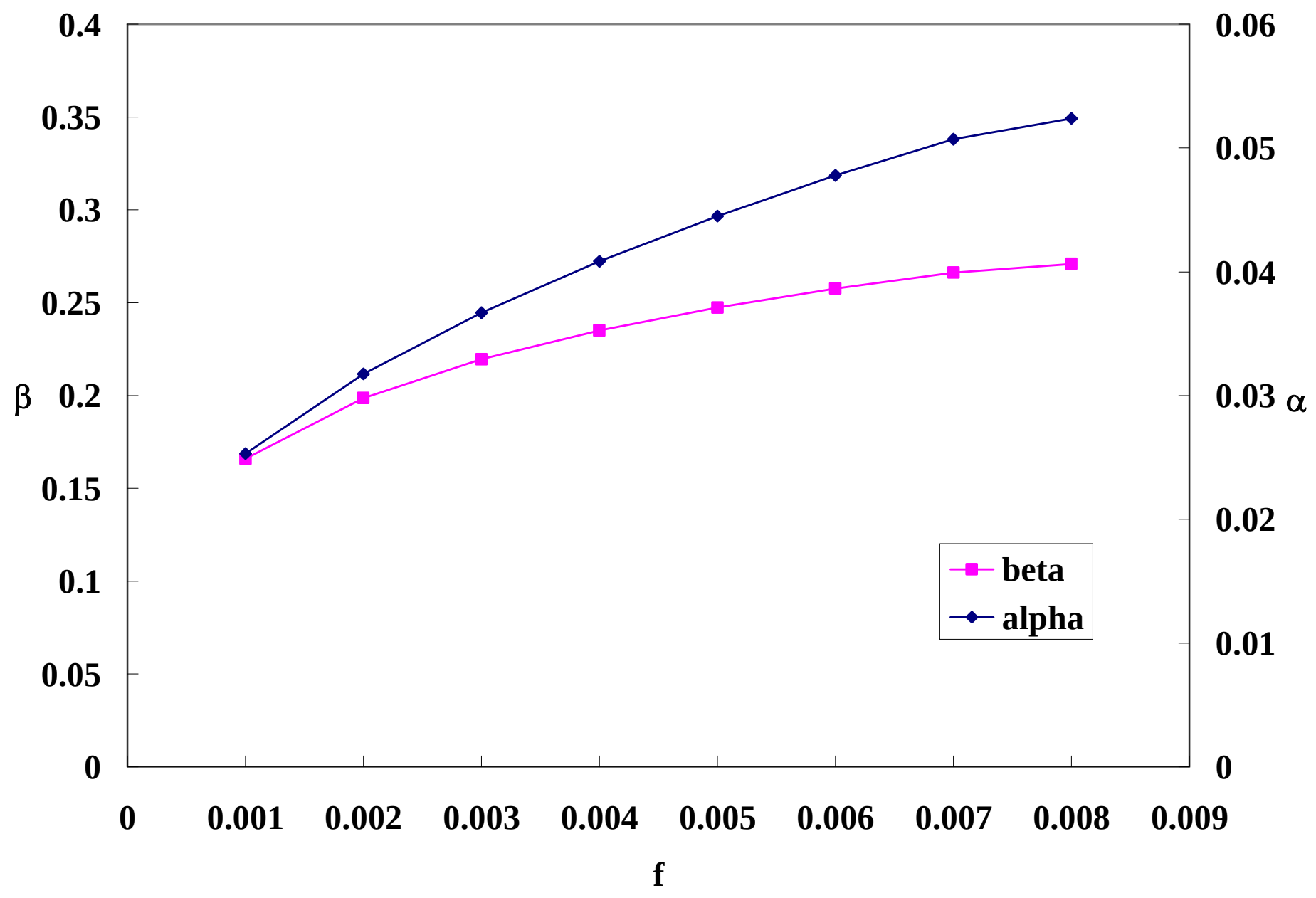

Figure 1. Variations of factors $\alpha$ and $\beta$ with the friction factor.

Back substituting into equation (7) gives

$$f_x = \frac{\alpha^{\frac{1}{\beta+1}}\left[(\beta+2)(\beta+1)\right]^{\frac{-\beta}{\beta+1}}\left[\beta(2-\beta)\right]^{\frac{\beta}{\beta+1}}}{\mathrm{Re}_x^{\frac{\beta}{\beta+1}}} = \frac{\alpha'}{\mathrm{Re}_x^{\frac{\beta}{\beta+1}}} \tag{34}$$

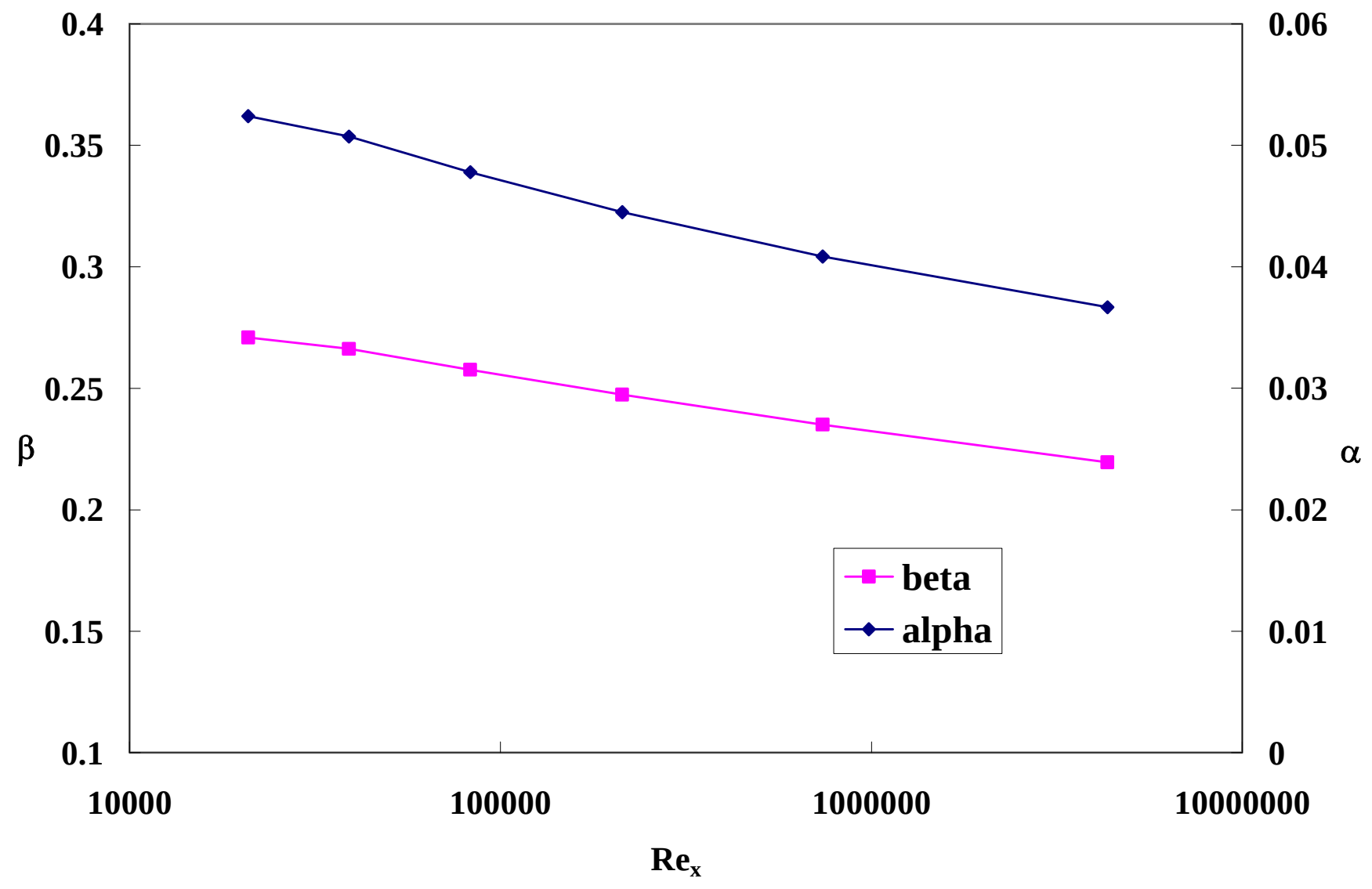

Figure 2. Variations of factors $\alpha$ and $\beta$ with the Reynolds number and friction factor.

The variations of $\alpha$ and $\beta$ with the Reynolds number are shown in Figure 2. Equation (34) compares very well with the empirical formulation of Nikuradse (Schlichting, 1960 p.600)

$$f_x = \frac{0.0592}{\mathrm{Re}_x^{1/5}} \tag{35}$$

A logarithmic correlation is easily obtained by substituting equation (17) into (20)

$$\frac{1}{\sqrt{f_x}} = 4.07 \ln\left( \mathrm{Re}_x^{\frac{1}{\beta+1}} \; f_x^{\frac{1}{2}} \left[ \frac{\alpha(\beta+1)(\beta+2)}{2(2-\beta)} \right]^{\frac{1}{\beta+1}} \right) + 3.276 \tag{36}$$

or alternately

$$\frac{1}{\sqrt{f_x}} = 4.07 \ln\left( x^+ \; \mathrm{Re}_x^{\frac{-\beta}{\beta+1}} \right) + \frac{4.07}{\beta+1} \ln\left[ \frac{\alpha(\beta+1)(\beta+2)}{2(2-\beta)} \right] + 3.276 \tag{37}$$

A formula for the average friction factor over a plate of length $L$ is more difficult to obtain formally because the factors $\alpha$ and $\beta$ in equations (34) and (36) are also a function of $\mathrm{Re}_x$, unlike the fixed average values taken by Nikuradse in equation (35). We can either integrate numerically equations (34) and (36) over the distance $L$ or circumvent the issue approximately by treating $\alpha$ and $\beta$ as constants over sections. Then equation (34) becomes

$$f_L = \frac{\alpha''(\beta''+1)}{\mathrm{Re}_L^{\frac{\beta''}{\beta''+1}}} \tag{38}$$

where $\alpha''$ and $\beta''$ are the log mean average values of $\alpha'$ and $\beta$ over the range of Reynolds number up to $\mathrm{Re}_L$.

# 3 Verification and discussion

Equation (38) correlates well the data reported in Schlichting' book (1960, p 600).

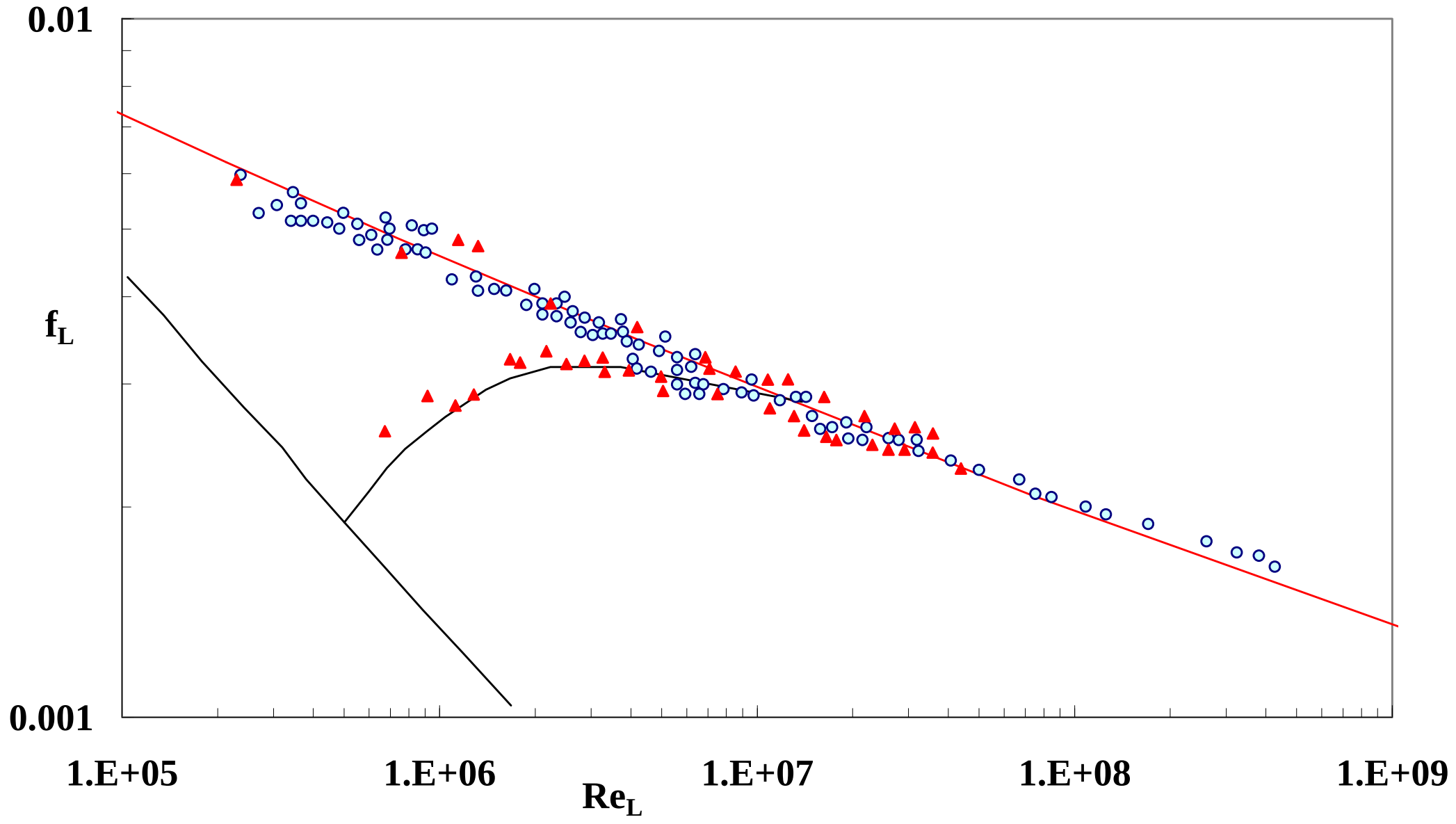

Figure 3. Average friction factor for turbulent boundary layer flow past a flat plate predicted by equation **Error! Reference source not found.**).

The present analysis shows that a Blasius type power law correlation for the friction factor does not need to empirical; it can be derived from fundamental theoretical considerations. It can be applied to situations other than flat plates. An example has been presented for non-Newtonian pipe flow (Trinh, 1993). The present technique also gives a closed solution for quick estimates of the friction factor in external boundary layers which can be useful in optimisation of equipment designs. More importantly, it gives us confidence in the use the power law velocity profile as a fundamental legitimate description of velocity profiles in turbulent flow.

As discussed in previous postings (Trinh, 2009), the most distinctive features of turbulent flows are the ejections (Kline et al., 1967) that can viewed as intermittent jets of wall fluid in cross flow. Many authors e.g. Chassaing et al. (1974), Camussi et al.(2002), have found that the path of the jet in the far field region follows a power function of the form

$$\frac{y}{D} = A\left(\frac{x}{D}\right)^n \tag{39}$$

The portion near the wall is quasi-linear and it slope correlates quite well with Karman's constant (Trinh, 2010b). Experimental confirmation was given by Chen and Blackwelder (1978) who followed these large scale motions by using temperature contamination from a heated wall. They observed a sharp temperature front associated with the upstream side of the turbulent bulges extending across the entire log-law region and related to the bursts from the wall. The matching of the semi-logarithmic and power law descriptions of the velocity profile rests on this physical visualisation.

# 4 Conclusion

The Blasius power law correlation for the friction factor has been derived theoretically. The coefficients in this correlation are shown to be dependent on the Reynolds number. Published experimental data is well correlated.